\documentclass[useAMS,usenatbib]{mn2e}
\usepackage{txfonts}
\usepackage{graphicx}
\usepackage{natbib}

\def\del#1{{}}

\title[Camouflaged galactic CMB foregrounds: total and polarized
  contributions of the kSZ effect]{Camouflaged galactic CMB foregrounds: total and polarized
  contributions of the kinetic Sunyaev Zeldovich
  effect}
\author[Andr{\'e} Waelkens, Matteo Maturi \& Torsten En\ss lin]
       {Andr{\'e} Waelkens$^{1}$, Matteo Maturi$^{1,2}$, Torsten En\ss lin$^{1}$ \\
$^{1}$Max Planck Institut f{\" u}r Astrophysik, Garching 85741, Germany\\
$^{2}$Institut f{\"u}r Theoretische Astrophysik, Heidelberg 69120, Germany\\}
\begin{document}
\date{Accepted 20?? December ??. Received 20?? December ??; in original form 20?? October ??}

\pagerange{\pageref{firstpage}--\pageref{lastpage}} \pubyear{2002}

\maketitle

\label{firstpage}

\begin{abstract}
We consider the role of the galactic kinetic Sunyaev Zeldovich (SZ) effect as a CMB
foreground. While the galactic thermal Sunyaev Zeldovich effect has
previously been studied and discarded as a potential CMB foreground, we find
that the kinetic SZ effect is dominant in the galactic case. We analyse the
detectability of the kinetic SZ effect by means of an optimally matched filter
technique applied to a simulation of an ideal observation. We obtain no detection, getting a
S/N ratio of 0.1, thereby demonstrating that the kinetic SZ effect can also
safely be ignored as a CMB foreground. However we provide maps of the expected
signal for inclusion in future high precision data processing. Furthermore, we
rule out the significant contamination of the polarised CMB signal by second
scattering of galactic kinetic Sunyaev-Zeldovich photons, since we show that the scattering of the
CMB quadrupole photons by galactic electrons is a stronger effect
than the Sunyaev Zeldovich second scattering, and has already been shown \del{by
\citeauthor{2005PhRvD..71f3531H}} to produce no significant polarised
contamination. We confirm the latter  assessment also by means of an optimally
matched filter.
\end{abstract}


\section{Introduction}
The role of the Sunyaev Zeldovich effect (hereafter SZ) as a CMB foreground has
  been widely studied for clusters of galaxies (see for example
  \cite{1972CoASP...4..173S}, \cite{1980ARA&A..18..537S},
  \cite{1999PhR...310...97B},  \cite{2005MNRAS.363...29D} and
  \cite{2006MNRAS.370.1713S}). Due to the high cluster temperatures the
  relevant effect in these cases is the thermal SZ effect (hereafter tSZ), while the kinetic
  SZ effect (hereafter kSZ) is secondary.
In this work we study the galactic SZ effect. The relevance of the tSZ effect
  as a galactic foreground has been ruled out by previous studies \cite[see
  e.g.][also section \ref{sec:tSZestimate}]{2007ApJS..170..288H}. As we will
  show, however, the kSZ effect makes a significantly higher contribution. This
  is confirmed by an independent galactic kinetic SZ effect simulation made by \citet{2007arXiv0705.3245H}, with whose results ours agree.  Even though the galactic kSZ effect is more significant than the
  tSZ effect, we demonstrate by means of the optimally matched filter
  technique that it is still not a significant CMB foreground.

We extend our search by also investigating whether the second scattering of 
galactic kSZ effect photons could present a significant CMB polarisation
foreground. 
The strength of the the polarised signal due to Thomson scattering
by galactic electrons is determined by the intensity of the incident quadrupole
radiation in the rest frame of the electron. Hence we assess the importance of
a polarised foreground by the intensity of the incident quadrupole radiation source.

We will show, that the polarisation signal due to the
intrinsic CMB quadrupole is far more significant than
that caused by secondary scattering of the kSZ induced quadrupole. In addition
we also examined the relevance of the CMB quadrupole induced by the mildly relativistic motion of the galaxy in the CMB
rest-frame. This effect, however, is known to be an order of magnitude smaller than the
intrinsic CMB quadrupole \citep{2007ApJS..170..288H} hence, although it is
slightly more relevant for polarisation production than the kSZ quadrupole,
also negligible. The CMB quadrupole induced polarisation by galactic electrons
has been shown to be an insignificant CMB polarisation foreground
\citep{2005PhRvD..71f3531H}. We confirm this result in sec. \ref{sec::PolFilt},
where we demonstrate by means of a matched filter that no galactic signal can
be detected even assuming an idealised experiment. Hence polarisation
contributions due to the weaker galactic kSZ effect can safely be ignored as CMB
polarisation foreground. 

The structure of the paper is the following.
In section \ref{kSZtheory} we give a brief description of the SZ
effect, while in section \ref{QUestimate} we describe the Thomson scattering theory
and check for the relevance of the kSZ effect polarisation signal. Section
\ref{kSZgal} describes our simulation of the galactic SZ effect and the polarised Thomson scattering
emission. In Section \ref{filtering} we describe our attempt to measure the SZ
and polarised Thomson scattering signature in our simulations. We draw conclusions in Section \ref{sec::Con}
and discuss how both the kSZ and the polarised Thomson
signal can be ignored as CMB foregrounds. 

\section[]{The Sunyaev-Zeldovich effect}
Here we present a brief theoretical introduction to the SZ effect and perform an
order-of-magnitude estimate of the galactic tSZ and the kSZ
effects as a first step toward investigating their importance as CMB foregrounds. We
do also check whether any relevant contamination of the CMB polarisation
signal due to the galactic kSZ effect is expected.

\subsection{Basics}
\label{kSZtheory}
The SZ effect is a distortion of the CMB signal
caused by the scattering of the CMB photons on moving free electrons
\citep[see ][]{1972CoASP...4..173S, 1980ARA&A..18..537S}. 
There are two components of the SZ effect: the thermal SZ effect due to the
electron thermal velocities and the kinetic SZ effect caused by the
bulk motion of electrons in the CMB rest frame.
The CMB spectrum distorted by the Compton scattering is given by
\begin{eqnarray}
\label{eq:Inu}
I(x)&=&I_{CMB}(x)+\delta I_{tSZ}(x)+\delta I_{kSZ}(x)\\
&=&i_0 [i(x)+g(x) y-h(x) \overline{\tau \beta}] , \nonumber
\end{eqnarray}
where $x=h\nu/(kT)$ is the dimensionless frequency. The CMB blackbody
spectrum is
\begin{equation}
\label{bbody}
I_{CMB}(x)=i_0 \, i(x) ,
\end{equation}
with normalisation factor $i_0=2(kT_{cmb})^3/(hc)^2$ and spectral shape $i(x) = {x^3}/ (e^x-1)$.
The second term on the right hand side of Eq. \ref{eq:Inu}
describes the tSZ effect. This effect has a
spectral shape given by
\begin{eqnarray}
g(x)=\frac{x^4 e^x}{(e^x-1)^2} \left ( x \frac{e^x+1}{e^x-1} -4 \right )\; , 
\end{eqnarray}
and amplitude given by the Comptonization parameter for a given
line-of-sight (LOS),
\begin{eqnarray}
\label{eq:compton_y}
y=\frac{\sigma_T}{m_e c^2} \int dl \, n_e(l) \, kT_e(l) .
\end{eqnarray}
The last term on the right hand side of Eq.  \ref{eq:Inu}
represents the kSZ effect. It has the spectral shape
\begin{eqnarray}
h(x)=\frac {x^4 e^x} {(e^x-1)^2}
\end{eqnarray}
and depends on the LOS integral
\begin{eqnarray}
\label{theo:tb}
\overline{\tau\beta}=\sigma_T \int dl \,  n_e(l) \, \beta (l) .
\end{eqnarray}
Here $\beta$ is the velocity component along the LOS in units of the speed of light.

It is common practice to display CMB measurements in terms of temperature
distortion maps. This is specially convenient for the kSZ effect, which can be
interpreted as a frequency independent change to the original CMB temperature.
The spectral distortions introduced by the kSZ effect are
\begin{eqnarray}
\label{eq:kSZapprox}
I(x')\simeq I(x)+ \left(\frac{\partial I}{\partial x'}\right)_{x'=x} \!\!\!\!\!\! (x'-x)=i_0 \left
(i(x)-h(x)\frac{\delta T}{T_{CMB}} \right )\; ,
\end{eqnarray} and therefore identical to spectral distortions generated by a
temperature change, to first order. Here $x'=\frac{h \nu}{kT'}$, where
$T'=T_{CMB}+\delta T$ and $\delta T \ll T_{CMB}$.
Comparing Eq. \ref{eq:Inu} and \ref{eq:kSZapprox} suggest the identification $\overline{\tau \beta} = \delta T_{kSZ}/ T_{CMB}$.
In contrast, the tSZ spectral
distortions cannot be described by a simple frequency-independent temperature change. However, for
our galaxy the kSZ effect dominates by orders of magnitude over the tSZ effect, as we show in the following.

\subsubsection{Order of magnitude estimate of the galactic thermal SZ effect}
\label{sec:tSZestimate}
Previous works have ruled out the relevance of the thermal SZ effect as a
foreground for the CMB \citep[see ][]{2007ApJS..170..288H}. They estimate
$y\simeq 2\times 10^{-8}$ by assuming a maximal temperature, density and LOS
distance in Eq. \ref{eq:compton_y}. Thus they argue that the SZ effect can
safely be ignored as a diffuse contaminating foreground signal. Indeed the
galactic tSZ effect would produce a variation in temperature about three orders
of magnitude smaller than the CMB primary anisotropies, at most.

\subsubsection{Order of magnitude estimate of the galactic kinematic SZ effect}
\label{sec:kSZestimate}
We will now attempt an estimate on the kSZ signal strength. 
The variation in temperature due to the kSZ (in the CMB rest-frame) is given by $\delta
T/T_{CMB}=\overline{\tau \beta}$ which is a function of the thermal electron density and the
LOS components of the electron velocities in the CMB rest-frame
(Eq. \ref{theo:tb}). As a rough estimate, we consider a homogeneous gas
distribution with electron density $n_e\sim0.1 {\rm cm}^{-3}$
\citep{2002astro.ph..7156C}, and a constant velocity of $371 {\rm km/s}$
\citep[see][]{1996ApJ...473..576F} all along the LOS, with a length $L=20 \, {\rm
  kpc}$, through the galactic plane.
Hence, 
\begin{eqnarray}
\label{kSZestimate}
\overline{\tau \beta} = \delta T_{kSZ}/T_{CMB} \sim \sigma_T \, L \, n_e \, \beta \sim
4\cdot 10^{-6} \nonumber .
\end{eqnarray}

This toy scenario portrays how large the kSZ effect
amplitude could maximally be in the direction of the galactic plane. Though still smaller
than the CMB anisotropy signal, it is significantly greater than the thermal SZ
effect (about two orders of magnitude). This motivates us to undertake a
more precise analysis of this phenomenon by
constructing a synthetic map of the galactic kSZ effect (see
sec. \ref{sec::kSZmapConst}) and to check whether it can be a
non-negligible CMB foreground (see sec. \ref{filtering}). 

\subsection{Polarisation due to Thomson scattering} \label{QUestimate}
Polarisation due to Thomson scattering depends only on the quadrupole of the
incoming radiation as is briefly demonstrated in the following.

The Q and U Stokes parameters due to Thomson scattering for each LOS are
computed for a local coordinate system whose xy-plane is orthogonal to the LOS
and whose z-direction points to the observer. To consistently compute the Stokes parameters for all directions on the sky we adopt the convention where the x-axis points in
negative $\hat \phi '$ and the y-axis in negative $\hat \theta '$ direction of
the observer-centric coordinate system whose $\rm{x'}$ axis points towards the
galactic centre while the $\rm{z'}$ axis points towards the galactic north
pole.
Assuming the usual $\theta$ and $\phi$ angles for the local coordinate
system we have 
\begin{eqnarray}
\label{Q}
Q=\frac{3 \sigma_T}{16 \pi} \int dl \, n_e(l)  \int d\Omega \, \sin^2\theta \,
\cos 2\phi \,
I(\theta, \phi, l) , \\
U=\frac{3 \sigma_T}{16 \pi} \int dl \,  n_e(l) \int d\Omega \, \sin^2\theta \,
\sin 2\phi \, I(\theta, \phi, l) . \label{U}
\end{eqnarray}

The terms $\sin^2 \theta \, \cos 2 \phi=\sqrt{8\pi/15} \left (Y_2^2(\theta,
  \phi)+Y_2^{-2}(\theta, \phi) \right )$ and $\sin^2 \theta \, \sin 2
\phi=-i\sqrt{8\pi/15} \left (Y_2^2(\theta, \phi)-Y_2^{-2}(\theta, \phi) \right )$ are linear
combinations of l=2 spherical harmonics, i.e. the quadrupole (here spherical
  harmonics are denoted by $Y_l^m(\theta, \phi)$). Hence only the quadrupole part of
$I(\theta, \phi)$ survives the integration over the whole sphere:
\begin{eqnarray}
\label{intsphQ}
\int d\Omega \, \sin^2\theta \, \cos 2\phi \, I(\theta, \phi)& = & \sqrt{\frac{8\pi}{15}}\left( a_{22}+a_{2-2}
  \right) \nonumber \\
\label{intsphQ2}
& = & \sqrt{\frac{32\pi}{15}}{\rm Re} \, a_{22}
\end{eqnarray}
and similarly
\begin{eqnarray}
\int d\Omega \, \sin^2\theta \, \sin 2\phi \, I(\theta, \phi)& = & -i\sqrt{\frac{8\pi}{15}}\left( a_{22}-a_{2-2}
  \right) \nonumber \\
\label{intsphU2}
& = & \sqrt{\frac{32\pi}{15}}{\rm Im} \, a_{22} .
\end{eqnarray}
Here, the $a_{lm}$ are the spherical harmonic coefficients. We have made use of
the property $a_{l \, -m}=-1^m a_{lm}^*$ of the real valued brightness
distribution, with the asterix denoting complex conjugation.

\subsubsection{Estimating the relevance of the polarised signal}
Here, we asses the typical strength of the polarisation signal caused by
second scattering of the kSZ photons. To do so we compare it to
other possible sources of polarised radiation. In order to identify the
dominant polarisation signal we only have to find the strongest quadrupole
source. 

Table \ref{tableQuadrupole} lists the total power of all possibly relevant quadrupole sources. Notice that other foregrounds like synchrotron or dust
emission present even stronger quadrupoles, however, since they have
frequency dependent brightness temperatures, they can, in principle, be removed. 
We see that the CMB quadrupole is by far the most
significant contribution. The Doppler quadrupole due to Earths motion in the
CMB rest-frame is negligible, while the quadrupole due to second scattering of
the kSZ effect is even smaller.

We first perform an order of magnitude estimate
to obtain the upper limit for the polarised signal due to the scattering of the
CMB quadrupole by galactic thermal electrons:

We refer to the CMB power
spectrum given by \cite{2007ApJS..170..288H}. In their table 7, they display the quantity $\Delta
T_2^2=211 \mu K^2$. It is given by $\Delta T_l^2=l(l+1) C_l/(2\pi)$, and
$C_l=1/(2l+1)\sum_{m=-l}^l |{a_{lm}}|^2$. For a rough upper limit estimate, we assume all
quadrupole power to be concentrated on only the real part of $a_{22}=\sqrt{5
  C_2}$. 
Assuming a homogeneous thermal electron number density $n_e\sim0.1 \, 
{\rm cm^{-3}}$  \citep[again following ][ as in section \ref{sec:kSZestimate}]{2002astro.ph..7156C} along a LOS of $L\sim20 \, {\rm kpc}$ (say towards the galactic centre) we get,
\begin{eqnarray}
Q &=&\sigma_T \sqrt{\frac{3}{40 \pi}} L \, n_e \, {\rm Re} \, a_{22} \simeq
2\cdot 10^{-2} \, {\rm \mu K} .
\end{eqnarray}
The polarised CMB intrinsic anisotropies have maximum amplitudes of $\sim 10 \%$ of
the total CMB anisotropies whose maxima are typically around a few ${\rm 100
  \mu K}\;$ \citep[see figure 15 of][]{2007ApJS..170..335P}. Hence, the
polarisation signal due to Thomson scattering on the galactic electrons from
the CMB radiation (only a few $10^{-2}\,{\rm \mu K}$) is likely to be undetectable
on any Q or U maps. The much weaker kSZ contribution (see table
\ref{tableQuadrupole}) is thus likely to be an insignificant contamination
to the CMB polarisation. 
This is supported by detailed computations of the Thomson scattering of
the CMB quadrupole by galactic electrons performed by
\cite{2005PhRvD..71f3531H}. The authors verify that the galactic Thomson signal
is not a significant contamination to the intrinsic CMB polarisation.

Complementary to the arguments presented in \cite{2005PhRvD..71f3531H}, in
sec. \ref{sec::PolFilt} we confirm by means of the matched filter technique
applied to our simulations that the signal is not detectable, consequently
neither is the one due to the second scattering of kSZ photons.

\begin{table}
\begin{tabular}{|lr|c|} \hline
Quadrupole source  & $C_2$ \\ \hline 
CMB    & $221 {\rm \mu K^2}$ \\ 
relat. Doppler effect & $3.45 {\rm \mu K^2}$ \\ 
kSZ effect & $10^{-2} {\rm \mu K^2}$ \\ 
\end{tabular}
\caption{\label{tableQuadrupole} Amplitude of the quadrupole moments in order
  of their significance. The relativistic Doppler effect quadrupole moment is
  calculated for $\beta \sim 1.2 \cdot 10^{-3}$. The CMB quadrupole given by \citet{2007ApJS..170..288H} includes this value (i.e. they do not
  attempt to subtract it from the measured value). The kSZ quadrupole moment is
  obtained from our simulation in sec. \ref{sec::kSZmapConst}. Other
  foregrounds with frequency dependent brightness temperatures like synchrotron
  or dust emission are assumed to be removed.}
\end{table}

\section{Simulations}
\label{kSZgal}
\subsection{The code}
We modified the Hammurabi-code originally designed for simulating
galactic synchrotron emission \citep{Hammurabi, 2006AN....327..626E}, to calculate the
LOS integral given in Eq. \ref{theo:tb} (for the kSZ effect) and in
Eq. \ref{Q} and \ref{U} (for the polarisation due to Thomson scattering). This
code uses the HEALPix pixelisation scheme \citep{2005ApJ...622..759G}.
\subsubsection{Line-of-sight integration for the kSZ effect}
The flux associated with an observation beam, whose width and shape is related
to the HEALPix pixels, is calculated by taking samples of the thermal electron
density and the velocity along the LOS at equal radial intervals. To compensate
for the widening of the beam cross section as one moves farther away from the
observer, the observation beam is, after a certain distance, split into 4
sub-beams over which we then average. Each sub-beam is again individually
sampled at the same regular radial intervals as before, and split again after
twice the first distance, and so forth. This avoids an under-sampling
of the thermal electron density and LOS velocity at large distances from the
observer. 

\subsubsection{Line-of-sight integration for the polarised galactic Thomson scattering}
In this case only the electron density is integrated along the line of
sight, however the computation is performed in an analogous manner to the previous paragraph. The
resulting map is multiplied by the $a_{22}$ component after a rotational
coordinate transformation for each different LOS in
order to get the respective $Q$ (Eq. \ref{Q}) and $U$ (Eq. \ref{U}) Stokes
parameters. The computation of the proper $a_{22}$ coefficients is described in appendix \ref{appQU}.

\subsubsection{Galaxy model}
\label{sec:GalMod}
Two input models are needed, a model for the thermal electron
distribution and a model for the galactic rotation curve.  
\begin{enumerate}
\item
  For the electrons we use the NE2001 model \citep[see
  ][]{2002astro.ph..7156C}. This model subdivides the galaxy in several large
  scale structure elements (thin disk, thick disk, spiral arms,...) as well as
  some local small scale elements such as supernovae bubbles. 

\item
  For the galactic rotation curve
  we assume a constant value of $220 \, {\rm km/s}$ above a galactocentric
  distance of $2 \, {\rm
  kpc}$. Below this distance the velocity falls off linearly to zero while approaching the
galactic centre. This is an approximation for the observed rotation
curve \citep[see e.g.][and references therein]{2002ApJ...573..597K}, which
should be adequate to determine whether the galactic kSZ effect can be
detected.
\end{enumerate}
\subsection{Constructing the kSZ map}
\label{sec::kSZmapConst}
The velocity of each point of the galaxy with respect to the CMB rest-frame is
subdivided into two components. The rotation velocity of the galaxy ($\beta_{rot}$) and the
position-independent translation velocity of the galaxy ($\beta_{trn}$). The
first quantity is obtained from our rotation curve model. The later quantity is
obtained by subtracting our solar systems galactic rotation velocity vector
($220 \,  {\rm
  km/s}$ in $l=\pi/2$ direction) from the measured total CMB frame velocity
\citep[$371 {\rm km/s}$ in $l\simeq 1.47 \pi$ and $b\simeq 0.27 \pi$ direction, 
see][]{1996ApJ...473..576F}. Since the resulting $\beta$ is small($\sim
10^{-3}$), we can safely ignore relativistic effects when adding velocities. 

However, relativistic effects may need to be considered for the temperature
transformation.  So far we obtained the $\overline{\tau \beta} = \delta
  T/T_{CMB}$ sky in the CMB rest-frame. The relativistic temperature
transformation is given by $\delta T_{obsf}=\delta T_{cmbf} \left [\gamma \left
    (1+(\beta_{tot} \, \mu_{obsf}) \right) \right]^{-1}$ .
Here $\mu_{obsf}$ is the cosine of the angle between the opposite of the CMB
frame motion direction and the LOS in the observer rest-frame and $\gamma=(1-\beta^2)^{-1/2}$.
The relativistic correction to the temperature is small, in the extreme a
factor $\sim (1\pm10^{-3})$, which can be safely ignored, leaving $\delta
T_{obsf}=\delta T_{CMBf}$. Thus to obtain the kSZ map we can neglect all
relativistic effects.

\begin{figure}
\resizebox{\hsize}{!}{\includegraphics{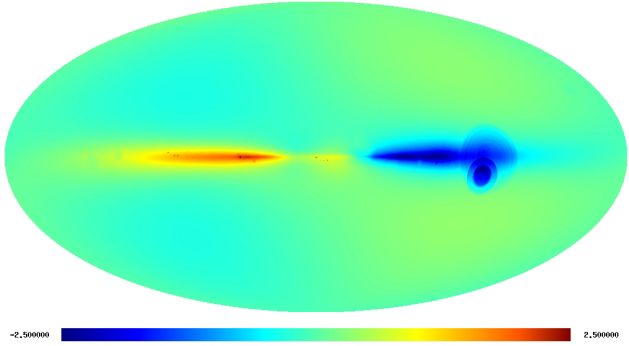}}
\caption{\label{fig:tbmapLISM} $\delta T=\overline{\tau \beta} \cdot T$ map in ${\rm
   \mu K}$. The monopole and dipole components of the map have been
  subtracted in analogy to what typically is done for the observations (e.g. WMAP). Our  simulation shows that the maximum amplitude of the map
  decreases by about 20$\%$ once the kSZ dipole is subtracted. Notice the
  features due to over dense or under dense regions in the local interstellar
  medium.}
\end{figure}

Finally, we obtain the kSZ map by inserting our galaxy model in
Eq. \ref{theo:tb}. The resulting map is shown in figure
\ref{fig:tbmapLISM}. Note the local interstellar medium features of the thermal
electron density model (see sec. \ref{sec:GalMod}). We subtracted the monopole
and dipole in accordance with the standard practice of subtracting the monopole
and the relativistic-Doppler-effect induced dipole from maps displaying the CMB
anisotropies. Our simulations show that this has a small effect, decreasing the
maximum amplitude of the map by about 20$\%$.

The maximum amplitudes of the simulated signal are of the order of
${\rm \mu K}$, in accordance with the estimate obtained in section
\ref{sec:kSZestimate}, and two orders of magnitude smaller than the CMB intrinsic anisotropies.

\subsection{Constructing the polarised Thomson scattering maps}
\label{sec::constPolTho}
\begin{figure}
\resizebox{\hsize}{!}{\includegraphics{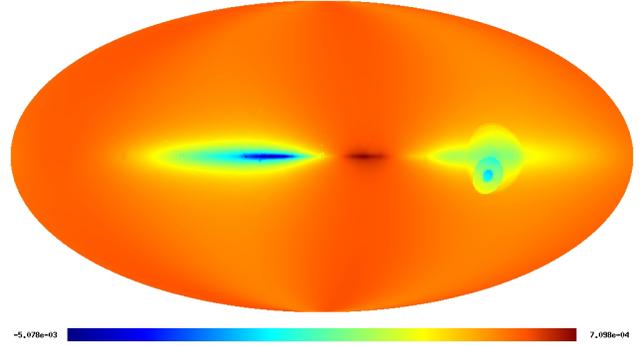}}
\caption{\label{Qmap} Stokes Q in $\rm {\mu K}$.}
\end{figure}

\begin{figure}
\resizebox{\hsize}{!}{\includegraphics{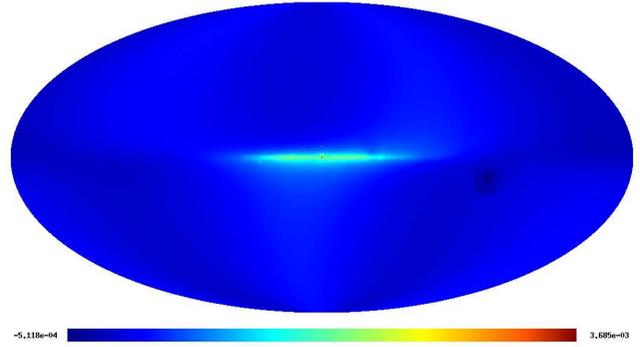}}
\caption{\label{Umap} Stokes U in $\rm {\mu K}$.}
\end{figure}

The expressions for the Q and U stokes parameters (Eqs. \ref{Q} and \ref{U})
assume that the negative z-axis of the coordinate system coincides with the LOS
direction. Thus it is necessary to compute the appropriate $a_{22}$ coefficient for
each direction on the sky. This is done by rotating a set of $a_{2m}$
coefficients, which are given in a certain coordinate system e.g. that in which
the WMAP data are presented, to the proper coordinate system in which the
negative z-axis is aligned with the LOS.
The procedure for rotating $a_{lm}$ coefficients is described in detail in
\citet{1988qtam..book.....V} and \citet{2004PhRvD..69f3516D}. For completeness,
we reproduce our special case in the appendix \ref{appQU}. 
Having performed the rotation we may apply Eqs. \ref{intsphQ2} and
\ref{intsphU2} to each position on the sky, which after performing a LOS integral of the thermal
electron density, finally allow us to compute Q and U, using Eqs. \ref{Q} and
\ref{U} respectively. The resulting full sky maps are shown in figures \ref{Qmap} and
\ref{Umap}. The maximum amplitudes are $\sim 10^{-3} \, {\rm \mu K}$, about
one order of magnitude lower than our estimate in section
\ref{QUestimate}. Out of the Q and U maps, the E and B mode spherical harmonic
coefficients are obtained. Since most of the CMB polarisation power is in E
modes, the B-modes signal of the polarised galactic Thomson scattering is more relevant as a CMB polarisation contamination than the corresponding
E-mode signal. In sec. \ref{sec::PolFilt} we search for B-mode signal
contamination by polarised galactic Thomson-scattering. 

\section{Filtering}
\label{filtering}
In order to estimate the feasibility of a detection of the effects discussed
above, we construct and apply an optimally matched filter to simulated data,
where we probe for the galactic kSZ and galactic polarised Thomson scattering
signatures.

The optimal matched filter was proposed for measuring radial bulk motion of
galaxy clusters through their kSZ imprint \citep{1996MNRAS.279..545H} and
subsequently found a wide range of applications in the literature, such as weak
lensing and the RS
effect \citep[see e.g. ][]{2005A&A...442..851M, 2006astro.ph..2539M}. A scalar adaptive generalisation of the filter
on the sphere was done by \cite{2006MNRAS.370.1713S}, and later extended to allow non-azimuthally symmetric objects to be detected
\citep{2006astro.ph.12688M}. 
We use a similar approach to \cite{2006MNRAS.370.1713S}. However, instead of
convolving the CMB signal and the filter for various positions and
orientations, we need only perform one convolution since we already know the
position and orientation of our signal. Details are presented in the appendix
\ref{app:filter}.  
The estimate of the maximum amplitude of the targeted signal is given by the
convolution of the filter $\Psi$  with the signal $S$
\begin{eqnarray}
\overline{A} = \int d\Omega \, \Psi(\theta, \phi) S(\theta, \phi) ,
\end{eqnarray}
while the variance is 
\begin{eqnarray}
\label{Aerror}
\sigma^2 & = & <\overline{A}>^2-A^2  = \sum_l \sum_m C_l \left |\psi_{lm} \right |^2 .
\end{eqnarray}
Here $\psi_{lm}$ is the spherical harmonic coefficient of the filter. The
 angular power spectrum of the noise ($C_l$) is either, in the kSZ case, the
 temperature angular  power of the CMB plus instrumental Gaussian noise, or, in the polarised Thomson-scattering case, the B-mode angular
power spectrum plus corresponding Gaussian instrumental noise.  
The filter is defined such that $\sigma^2$ is minimal while the bias $b=\langle
 \overline{A} \rangle - A =0$ is zero.  

\begin{figure}
\resizebox{\hsize}{!}{\includegraphics{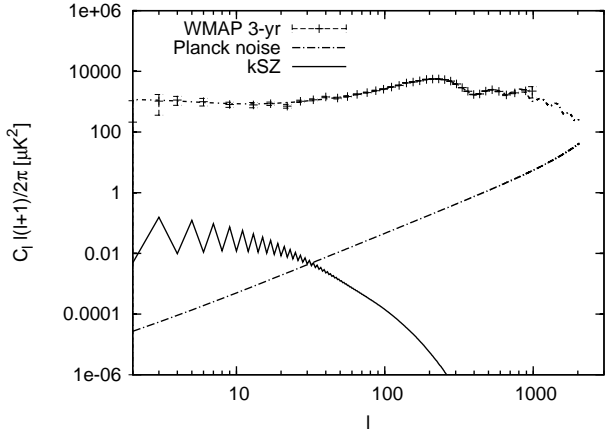}}
\caption{\label{fig:pspec} Comparing the CMB power spectrum to the simulated
  kSZ power spectrum. The saw like pattern for the kSZ effect power spectrum is
  due to the fact that even spherical harmonics are zero at the equator of the
  coordinate system, which in this case corresponds to the galactic
  plane. Hence odd modes will have more power.}
\end{figure}

\subsection{Filtering on simulated data: galactic kSZ}
\label{sec:FiltSim}
We consider the ideal situation in which the only additional contribution to
the CMB plus instrumental noise signal is the kSZ effect. 

\begin{figure}
\resizebox{\hsize}{!}{\includegraphics{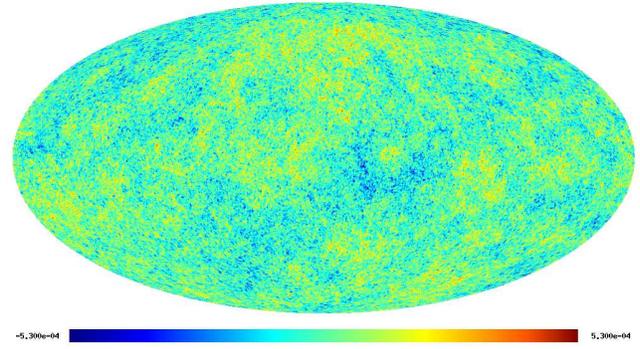}}
\caption{\label{fig:kSZplusCMB} Simulated CMB sky contaminated by the simulated galactic
  kSZ effect.}
\end{figure}

We do a simulation of the CMB sky with a sampling resolution of about $7 \,
{\rm arcmin}$ (corresponding to the HEALPix parameter NSIDE=512) and assume
experimental characteristics expected for a Planck-like experiment, i.e. a
resolution of 10 arcmin and a sensitivity of $5 \, {\rm \mu K /
  beam}$. Given that the structures of the simulated kSZ effect are mostly on
larger scales (see figure \ref{fig:pspec}), the sensitivity, instead of the
resolution, is the more relevant instrumental characteristic here. 
We generate a kSZ map with the same sampling resolution, add
it to the synthetic CMB sky and then convolve the map with a one degree
Gaussian kernel. This smooths the sharp borders of the simulation and also
ensures that our final map is properly sampled. The result
is shown in figure \ref{fig:kSZplusCMB}, in which no evidence of the kSZ effect
is visible.
\begin{figure}
\resizebox{\hsize}{!}{\includegraphics[angle=90]{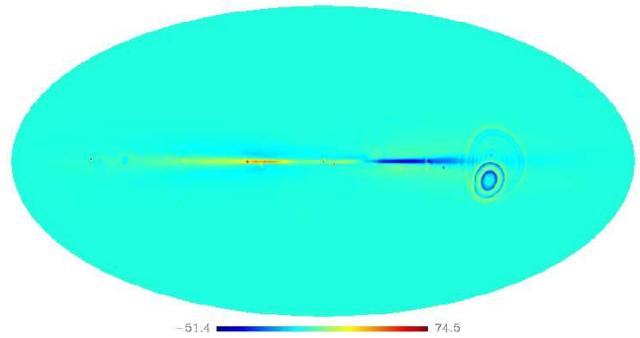}}
\caption{\label{fig:filter} The optimally matched galactic kSZ filter. Note the
  features due to the local inter stellar medium components of the thermal
  electron density model \citep{2002astro.ph..7156C}.}
\end{figure}

As a signal template we take the same convolved kSZ map previously used in defining the
data. This highly idealised approach (assuming a perfect signal template and no
contamination by other Galactic emission sources) is suitable to determine the
best possible filter performance.

The signal amplitude obtained by applying the filtering technique
is $-10 \pm 21 {\rm \mu K}$, with the variance given by
Eq. \ref{Aerror}. The expected maximum amplitude of the kSZ map, the quantity
we are estimating with the optimally matched filter, is $2.4 {\rm
  \mu K}$, according to our model. Hence we have a signal-to-noise ratio of 0.1. 

Although in this simulation no other foregrounds were present, no detection was
possible, demonstrating that the galactic kSZ effect can be ignored as a
CMB foreground.

\subsection{Filtering on simulated data: the B-mode signal}
\label{sec::PolFilt}
The measurement of CMB B modes is the next cosmological challenge and would
be very relevant for constraining inflationary models. The
CMB polarised power is overwhelmingly in the E-modes, contrary the polarised
galactic Thomson scattering. For this, our simulation shows a balance between
the E and B-mode power on large scales, and a difference by at most one order
of magnitude on smaller scales. This implies that
the galactic B-mode contamination of the CMB is relatively stronger than the corresponding
E-mode contamination. Therefore applying the same filtering technique as in
sec. \ref{filtering} we can investigate whether the polarised Thomson scattered
B-mode signal is detectable and, if so, whether it is a relevant CMB foreground.

We use the predicted CMB BB power spectrum $C^{BB}_l$ \citep[as given by the
CAMB code,][]{2000ApJ...538..473L} for a tensor-to-scalar
ratio of $10^{-5}$ \citep[following][]{2005PhRvD..71f3531H}, plus the
instrumental noise for a sensitivity of $7.6 \, {\rm \mu K / beam}$ and a resolution of $30$
arcmin (corresponding to polarisation instrumental characteristics of a Planck-like experiment), as the noise power spectrum in Eq. \ref{filter}. The
tensor spherical harmonic coefficients $b^{Thomson}_{lm} \rightarrow \tau_{lm}$
for the B modes of our polarised galactic Thomson-scattering simulation are
used as the spherical harmonic coefficients of our template.

This analogy to the previous case is permissible since $C^{BB}_l$ is completely
uncorrelated between each different mode (like the CMB temperature power
spectrum), and the CMB B mode signal is uncorrelated with the galactic B mode
signal.

We obtain an estimate of $-4 \cdot 10^{-3} \pm 5.8 \cdot 10^{-2} {\rm
  \mu K}$, against the simulated signal amplitude of $1.6 \cdot 10^{-3}
{\rm \mu K}$. Since the achieved signal-to-noise ratio is $\sim 0.03$, the Thomson
scattering by galactic thermal electrons is a negligible contribution to the
CMB polarisation signal. 

\section{Conclusions}
\label{sec::Con}
We have investigated the role of the galactic kinetic Sunyaev-Zeldovich effect as a CMB
foreground, both for the total and polarised emission. This was done by means of an optimally
matched filter based on a model of the signal and the
expected noise power spectrum. The filter returns an unbiased estimate of the
peak signal amplitude and is optimal in the sense that it provides the
minimum variance for such a measure. 

Regarding the non-polarised kinetic Sunyaev-Zeldovich emission as a
CMB foreground, we found that, although the kinetic Sunyaev-Zeldovich effect is
more relevant than the thermal Sunyaev-Zeldovich 
effect, it is still not strong enough to be detectable by means of an
optimally matched filter. 

Our simulation for the kinetic Sunyaev-Zeldovich effect shows the signal to be concentrated on the
galactic plane where the electron density is highest. The maximum amplitude of
the simulated signal (the quantity estimated by the filtering procedure) is
$2.4 {\rm \mu K}$. Assuming an instrument with resolution of 10 arcmin and a
sensitivity of $5 \, {\rm \mu K / beam}$, \del{where
foregrounds like dust or synchrotron radiation are supposed to have been
perfectly removed,} we obtain a
signal amplitude of $-10 \pm 21 {\rm \mu K}$. Hence, no kinetic
Sunyaev-Zeldovich effect signature could be detected.

We further demonstrate that the secondary scattering of kinetic Sunyaev-Zeldovich photons is negligible as a
polarisation source. It is a weaker effect than the
polarised emission due to Thomson scattering of CMB photons by galactic
electrons. Again by means of the optimally matched filter technique, we
determined that the latter effect is also irrelevant even as a CMB B-mode
foreground, assuming a tensor to scalar ratio of $10^{-5}$ (where most of the
polarised CMB power is in E-modes, and hence the B-modes are more easily
contamined). The polarised galactic Thomson-scattering signal is also
concentrated on the galactic plane. The maximum amplitude of the simulated signal is $8 \cdot 10^{-4}
{\rm \mu K}$. Assuming a Planck-like instrument with a resolution of 30 arcmin and
a sensitivity of $7.6 \, {\rm \mu K / beam}$, we estimate an amplitude of $-4
\cdot 10^{-3} \pm 5.8 \cdot 10^{-2} {\rm \mu K}$; thus no contamination of the CMB
B-mode signal is detectable.

In conclusion, galactic foregrounds due to the kinetic Sunyaev-Zeldovich effect are a certainly existing,
however perfectly camouflaged contamination of the CMB. Nevertheless, it is an expected physical effect which generates
1$\%$ temperature-fluctuation amplitude corrections on large scales, and thus should be taken into
account when estimating the error of precision measurements as the Planck
satellite will provide.

\section*{Acknowledgements}
AW would like to thank Bj{\"o}rn Malte Sch{\"a}fer, Martin
Reinecke, Amir Hajian, Christopher Hirata and Niayesh Afshordi for enlightening
discussions and comments. Also Stuart A. Sim, Tony Banday, Christoph Pfrommer,
Mona Frommert and Jens Jasche for corrections to the manuscript.

\bibliography{new}
\bibliographystyle{mn2e}

\appendix

\section[]{Filter derivation}
\label{app:filter}

In this section we describe how to obtain an optimally matched
filter function $\Psi(\theta, \phi)=\sum_l \sum_m \psi_{lm} Y_l^m(\theta,
\phi)$ for attempting to detect a non azimuthally-symmetric target signal (such
as the signal due to e.g. the galactic kSZ or the B mode due to galactic
Thomson scattering) in the presence of a Gaussian random noise (e.g. the sum of
CMB anisotropies, total or polarised, and the
instrumental noise). We used such a filtering procedure in
investigations of the RS effect \citep{RS_Matteo}.

Following the arguments layed out by \cite{1996MNRAS.279..545H}, \cite{2005A&A...442..851M} and \cite{2006MNRAS.370.1713S}, we assume the data to be
composed of the statistically homogeneous CMB anisotropies, some instrumental
Gaussian random noise and the targeted signal. 
The expansion of the data in spherical harmonics ($Y_l^m(\theta, \phi)$) is
\begin{eqnarray}
\label{eq:Sig}
S(\theta, \phi)= \sum_l \sum_m s_{lm} Y_l^m(\theta, \phi), 
\end{eqnarray}
where we model the complex coefficient $s_{lm}$ as
\begin{eqnarray}
\label{eq:SigComplex}
s_{lm}= A \tau_{lm}+a_{lm} .
\end{eqnarray}
We choose $A$ to be the maximum amplitude of the targeted
signal, while the morphology is encoded in the complex spherical harmonic
coefficient $\tau_{lm}$. The sum of the CMB and the noise sky is $a_{lm}=a'_{lm}+n_{lm}$,
where $n_{lm}$ is the instrumental noise modelled as a white Gaussian random
noise field. 
Note that the ensemble average $\langle a_{lm} \rangle=0$.

In the studies mentioned above, the convolution of the filter with the total
signal is performed for all positions on the sky (or the corresponding
analogous on the plane), in effect generating a convolution map described by
\begin{eqnarray}
\label{conv}
\Psi*S=\int d\Omega \, \Psi(\theta-\alpha, \phi-\beta) S(\theta, \phi).
\end{eqnarray}
In our case, however, we know the precise position and orientation of our target signal, therefore we need only to evaluate the convolution for
a single position on the sky. 
We define the estimate of the target signal strength to be
\begin{eqnarray}
\label{Aestimate}
\overline{A} = \int d\Omega \, S(\theta, \phi) \Psi(\theta, \phi) .
\end{eqnarray}
We want to estimate the real peak amplitude of
the target effect accurately. Hence we enforce an unbiased estimate 
\begin{eqnarray}
\label{zerobias}
b = A-<\overline{A}>=0 ,
\end{eqnarray}
with variance
\begin{eqnarray}
\label{Ap:error}
\sigma^2 & = & \langle \overline{A}^2 \rangle -A^2 = \sum_l \sum_m C_l \left |\psi_{lm} \right |^2 .
\end{eqnarray}
We enforce the condition that the variance be minimal by means of Lagrange
multipliers. We are looking for the filter $\Psi$ which would give the smallest
error for the estimate in Eq. \ref{Aestimate}, subject to the constraint of
being unbiased (Eq. \ref{zerobias}). Thus we solve
\begin{eqnarray}
\frac{\partial}{\partial \Psi^*} \left ( \sigma^2-\lambda b - \lambda' b^*
\right)=0 .
\end{eqnarray}
The resulting filter is given by
\begin{eqnarray}
\psi_{lm}=\frac{\lambda \tau_{lm}}{C_l} ,
\end{eqnarray}
which, when substituted in Eq. \ref{zerobias}, yields
\begin{eqnarray}
\lambda=\frac{1}{ \sum_l \sum_m \left(\left | \tau_{lm} \right |^2 /C_l \right
  )} \quad \mbox{and} 
\end{eqnarray} 
\begin{eqnarray}
\label{filter}
\psi_{lm}=\frac{1}{\sum_l \sum_m \left (\left | \tau_{lm} \right |^2 /C_l
  \right )} \frac{\tau_{lm}}{C_l} .
\end{eqnarray}
Analogous to the flat space results in \citet{1996MNRAS.279..545H}, the filter
gives preference to modes where the ratio of the target signal to the noise is largest.
The estimate of the target signal amplitude is performed using Eq.
\ref{Aestimate} with the error given by \ref{Ap:error} and the filter by
\ref{filter}.

\section{Rotating spherical harmonic coefficients}
\label{appQU}

Here we describe how a set of spherical harmonic quadrupole
coefficients ($a_{2m}$) transforms under a rotation of the coordinate system. In
particular we obtain the rotated spherical harmonic coefficient needed for
computing the Stokes Q and U parameters (Eqs. \ref{Q} and \ref{U} in
sec. \ref{QUestimate}).

The real-valued $a_{2m}$ coefficients for the CMB radiation given by
\cite{2007ApJS..170..288H}, are related to the complex-valued spherical
harmonics (the convention in this paper) as listed in table \ref{compxalm} .

\begin{table}
\label{compxalm}
\begin{tabular}{|lr|c|} \hline
$a_{2m}$  & complex value  \\ \hline 
$a_{2 \, 2}$   & $(-14.41-i18.8)/\sqrt{2}$ \\ 
$a_{2 \, 1}$   & $(-0.05+i4.86)/\sqrt{2}$ \\ 
$a_{2 \, 0}$   & $11.48$ \\ 
\end{tabular}
\caption{\label{table:complexAlm}  The $a_{lm}$ coefficients calculated from the real
  valued coefficients in table 6 of \citet{2007ApJS..170..288H}. These
  coefficients are obtained for the z-axis of the coordinate system pointing to
  the galactic north pole.}
\end{table}

The transformation of spherical harmonics under a rotation described by the
Euler angles $\alpha$, $\beta$ and $\gamma$ is given by \cite{1988qtam..book.....V}:
\begin{equation}
\hat Y_l^{m'}(\hat \theta, \hat \phi) = \sum_{m=-l}^l D_{mm'}^l (\alpha, \beta, \gamma) Y_l^{m}(\theta,
\phi) , \nonumber
\end{equation}
where $D^l_{mm'}(\alpha, \beta, \gamma)$ is the Wigner-D symbol. The hat
denotes quantities in the rotated coordinate system, otherwise quantities are
taken to be in the usual coordinate system where z points towards the galactic
north pole. In our case, the inverse transformation is needed,
\begin{eqnarray} {\label{eqYrot}}
Y_l^{m'}(\theta, \phi) 
&=&\sum_{m=-l}^l {D_{m'm}^l}^* (\alpha, \beta, \gamma) \hat Y_l^{m}(\hat\theta,
\hat\phi) .
\end{eqnarray}
The inverse of the Wigner-D symbol is given by 
\begin{eqnarray}
{D^{-1}}_{mm'}^l(\alpha, \beta, \gamma)={D_{m'm}^l}^*(\alpha, \beta, \gamma) ,
\end{eqnarray}
here the asterix denotes complex conjugation. 

We can write the total quadrupole intensity as a combination of spherical
harmonics in any coordinate system
\begin{eqnarray} {\label{eqItot}}
I_{l=2}(\theta, \phi) & = & \sum_{m=-2}^2 a_{2m} Y_2^m (\theta, \phi) = \sum_{m=-2}^2 \hat a_{2m} \hat Y_2^m (\hat \theta, \hat \phi) .
\end{eqnarray}

Substituting \ref{eqYrot} in \ref{eqItot}, multiplying by $Y_2^2(\theta, \phi)$, and
performing an integral over the whole sphere yields,
\begin{equation}
\hat a_{22}= \sum_{m'=-2}^2 a_{2m'} {D^2_{m'2}}^*(\alpha, \beta, \gamma) .
\end{equation}
The Wigner-D symbol is ${D^2_{m'2}}^*(\alpha, \beta, \gamma)=e^{im'\alpha}
{d^2_{m'2}}^*(\beta)e^im\gamma$, with
\begin{eqnarray}
{d^2_{m'2}}^*(\beta) \!\!\!\!\!\! & = & \!\!\!\!\!\! \left [ (2+m')! (2-m')! 4!\right ]^{1/2}
\frac{\cos^{(2+m')}(\beta/2)\sin^{(2-m')}(\beta/2)} {(2-m')!(2+m')!}
. \nonumber
\end{eqnarray}

We require the value of $\hat a_{22}$ in a coordinate
system which has its negative z-axis pointing in the direction specified by
the usual $\theta$ and $\phi$ angles. The corresponding
Euler angles are
\begin{eqnarray*}
\gamma=-\pi/2 \, , \; \beta=\pi-\theta \, , \; \alpha=\phi-\pi .
\end{eqnarray*}
Note that this means that the x-axis and y-axis point in negative $\phi$ and
negative $\theta$ directions, respectively. This is the coordinate system
in which the Stokes Q and U parameters are calculated in this work.

Inserting $\hat{a}_{22}$ in Eqs. \ref{intsphQ2} and \ref{intsphU2} and multiplying these
with the corresponding LOS integral of the thermal-electron density
and the appropriate constant pre-factor yields Eqs. \ref{Q} and \ref{U}. 

\bsp
\label{lastpage}

\end{document}